\documentclass[doublecol]{epl2}
\usepackage{graphicx}
\usepackage{amsfonts}
\usepackage{amsmath}
\usepackage{amssymb}
\usepackage{bm}
\usepackage{subfigure}
\usepackage{threeparttable}%
\usepackage[colorlinks=true ,citecolor=blue,urlcolor=blue]{hyperref}
\title{Detecting Majorana fermions by use of
superconductor-quantum Hall liquid junctions}
\author{Zheng-Wei Zuo\inst{1} \and Huijuan Li\inst{1} \and Liben Li\inst{1} \and L. Sheng\inst{2} \and R. Shen\inst{2} \and D. Y. Xing\inst{2}}
\shortauthor{Z.W. Zuo \etal}
\institute{
  \inst{1} School of Physics and Engineering, Henan University of Science and
Technology, Luoyang 471003, China\\
  \inst{2} National Laboratory of Solid State Microstructures, Department of
Physics, and Collaborative Innovation Center of Advanced Microstructures, Nanjing University, Nanjing 210093, China
}

\abstract{
The point contact tunnel junctions between a one-dimensional  topological superconductor
and  single-channel quantum Hall (QH) liquids are
investigated theoretically with bosonization technology and renormalization
group methods. For the $\nu=1$  integer QH liquid,
the universal low-energy tunneling transport is governed by the perfect Andreev
reflection fixed point with quantized zero-bias conductance
$G(0)=2e^{2}/h$, which can serve as a definitive fingerprint of the existence of a Majorana
fermion. For the $\nu =1/m$ Laughlin fractional QH
liquids, its transport is governed by the perfect normal
reflection fixed point with vanishing zero-bias conductance and bias-dependent conductance
$G(V) \sim V^{m-2}$. Our setup is within reach of the present experimental techniques.
}

\begin{document}

\maketitle

\section{Introduction}

Due to their potential applications in fault-tolerant topological quantum
computation~\cite{Nayak08RMP}, Majorana fermions have been of great interest
in condensed matter physics over the past
decade~\cite{Alicea12RPP,Beenakker12ARCMP,Franz15RMP}. Several solid-state
candidates for experimentally realizing topological superconductors (TSC) with
the Majorana fermions have been put forward, based on proximity coupling to
$s$-wave superconductors, such as the topological
insulators~\cite{FuL08PRL,FuL09PRB}, semiconductor quantum wires with
spin-orbit interaction~\cite{OregY10PRL,LutchynRM10PRL,StanescuTD13JPCM}, and
quasi one-dimensional (1D) chains of magnetic
atoms~\cite{Nadj-Perge13PRB,PientkaF13PRB,KlinovajaJ13PRL,BrauneckerB13PRL,VazifehMM13PRL}%
.

At present, verifying and detecting the existence of Majorana fermions remains
an outstanding experimental challenge~\cite{Franz15RMP}. The possible
experimental evidences of Majorana fermions have been shown in spin-orbit
coupled quantum wires subjected to a magnetic field and proximate to an
$s$-wave superconductor~\cite{Mourik12SCI,DasA12NTP,DengMT12NanoLett}. A
zero-bias conductance peak (ZBCP) was reported by these experiments. However,
other explanations of this anomaly ZBCP were also
suggested~\cite{LiuJ12PRL,BagretsD12PRL,PikulinDI12NJP}. How to distinguish
the ZBCP caused by the Majorana fermion from those by other mechanisms is an
important task. There are other proposals for detecting the Majorana fermions in TSC~\cite{PengY15PRL,GiladBS15PRB,JuanF14PRL,LiuD11PRB,BenjaminC10PRB}.

For a TSC coupled to other materials via point contact, the presence of
Majorana fermions can lead to many interesting transport
properties~\cite{LawKT09PRL,FidkowskiL12PRB,Affleck13JSM,KomijaniY14PRB,ClarkeDJ14NTP,VasseurR14PRX,LeeYW14PRB,ChaoSP15PRB,LutchynRM13PRB,PikulinDI15arXiv,ZuoZW16arXiv2}%
. When a normal metal lead, either noninteracting or interacting, couples to
Majorana fermions through electron tunneling, the Majorana fermions can induce
perfect Andreev reflections and may result in a quantized zero-bias
conductance $2e^{2}/h$ at low energies~\cite{LawKT09PRL,FidkowskiL12PRB}. For
the normal metal lead, various parameters such as the electronic interaction,
disorder, and finite length need to be taken into
account~\cite{FidkowskiL12PRB,Affleck13JSM,KomijaniY14PRB,ClarkeDJ14NTP,VasseurR14PRX,LeeYW14PRB,ChaoSP15PRB,LutchynRM13PRB,PikulinDI15arXiv,ZuoZW16arXiv2}%
, and they can result in a drastic departure from the noninteracting cases.
These factors have been analyzed by scattering matrix theory~\cite{LawKT09PRL}%
, Keldysh formalism~\cite{LutchynRM13PRB}, the Luttinger liquids theory and
renormalization group method~\cite{GogolinAO98Book,GiamarchiT04Book}. It is
highly desirable to overcome effects of these factors on the tunneling
spectrum. If the normal metal lead is replaced with a quantum Hall (QH)
liquid~\cite{Klitzing80PRL,Tsui82PRL,Laughlin83PRL}, the situation will be
quite different. The QH liquid is a topologically nontrivial system with
topologically-protected edge states. For the single-channel QH liquid lead,
the electronic interactions are suppressed or fixed, and effects of disorder
and finite length are also removed. No study to date, however, has examined
the advantages of using the QH liquid-TSC tunnel junctions. The integer
(fractional) QH states show precisely quantized plateaus at integer
(fractional) Hall conductance values of $e^{2}/h$. The effects of electronic
interactions are suppressed in the integer QH liquids. There are
fractionally-charged quasiparticles with fractional exchange statistics in
fractional QH states~\cite{WenXG92Review,WenXG95AP}. The integer and
fractional QH states support gapless edge excitations. Due to these intriguing
and exotic properties of the QH liquids, it is of both theoretical and
practical interest to investigate transport properties of the junctions
between the TSC and QH liquids.

In this work, we investigate the point contact tunnel junctions between a 1D
TSC and single-channel QH liquids with bosonization technology and
renormalization group methods. For the $\nu=1$ integer QH liquid, the perfect
Andreev reflection with quantized zero-bias conductance $2e^{2}/h$ at zero
temperature is predicted to occur, which is caused by the Majorana fermion
induced tunneling rather than by the conventional Cooper-pair tunneling. Such a
quantized conductance can serve as a definitive fingerprint of a Majorana
fermion, which is robust against details of the setup. On the other hand, for
the Laughlin fractional QH liquids ($\nu=1/m$ with $m$ the odd integral
greater than 1), the universal low-energy transport is governed by the fixed
point describing perfect normal reflection, which leads to a vanishing
zero-bias conductance and bias-dependent conductance of $G(V) \sim V^{m-2}$.
If the SC is topologically trivial, the bias-dependent conductance becomes
$G(V) \sim V^{4m-2}$. From these behaviors of $G(V)$, we can extract the
topological invariant $m$ controlled by the bulk fractional QH states.

\section{Theory and Discussion\label{theory}}

\begin{figure}[ptbh]
\centering{ \includegraphics[scale=1.3]{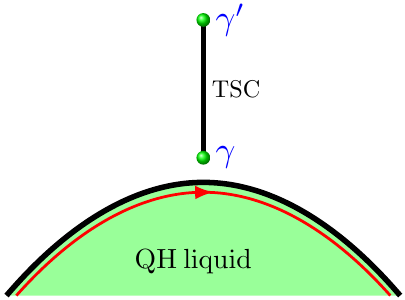}}\caption{(Color online)
Schematic illustration of the tunnel junction between a 1D TSC and a
single-channel QH liquid. The TSC is characterized by Majorana modes $\gamma$
and $\gamma^{\prime}$. The edge of the QH state can be described in terms of
chiral bosonic field $\phi$.}%
\label{Fig1}%
\end{figure}

The point contact tunnel junction we study is formed by a single-channel QH
liquid adjacent to a 1D TSC, as depicted in figure \ref{Fig1}. The 1D TSC can
be obtained by a spin-orbit coupled quantum wire subjected to a magnetic field
and proximate to an $s$-wave superconductor\cite{OregY10PRL,LutchynRM10PRL}.
The 1D TSC is characterized by the localized Majorana modes $\gamma$ and
$\gamma^{\prime}$ at end points. It is assumed that all the important energy
scales are smaller than the superconducting energy gap that is taken as the
unit of energy.

There are two types of single-channel QH liquid with filling fraction
$\nu=1/m$ where $m$ is an odd integer. One is the integer QH liquid with
$\nu=m=1$ and the other is the Laughlin fractional QH liquid with $m\geq3$.
Because of the high magnetic field, the electrons in the QH liquids are fully
spin-polarized (spin up). The model Hamiltonian for the whole system can be
written as%
\begin{equation}
H=H_{0}+H_{T}, \label{Htotal}%
\end{equation}
where $H_{0}$ is the edge Hamiltonian of the QH liquid and $H_{T}$ is the
tunneling Hamiltonian at the point contact. For the single-channel QH liquid,
the edge excitations are described by a single-channel chiral Fermi (or
Luttinger) liquid\cite{Halperin82PRB,WenXG95AP}. The edge Hamiltonian of QH
liquid is given by%

\begin{equation}
H_{0}=\frac{m\upsilon_{F}}{4\pi}\int dx\left(  \partial_{x}\phi\left(
x\right)  \right)  ^{2}, \label{Hzero}%
\end{equation}
with $\phi\left(x\right)$ a chiral boson field and $\upsilon_{F}$ the Fermi velocity.
The electron density operator is given by $\rho\left(  x\right)  =\partial
_{x}\phi\left(  x\right)  /2\pi$. The electron operators can be expressed as
\begin{equation}
\Psi_{\alpha}\left(  x\right)  =\Gamma_{\alpha}e^{im\phi\left(  x\right)  },
\end{equation}
with $\Gamma_{\alpha}$ the Klein factors which obey $\Gamma_{\alpha
}^{\dagger}=\Gamma_{\alpha}$ and $\left\{  \Gamma_{\alpha},\Gamma_{\beta
}\right\}  =2\delta_{\alpha\beta}$ to ensure the anti-commutation relations
between different fermion species. The Klein factors can be viewed as
additional Majorana fermions, which is important for studying related Majorana
fermion models~\cite{BeriB13PRL,Alexander13PRL}. Because of the
anti-commutation relations of electrons $\Psi_{\alpha}$ and Majorana fermion
$\gamma$, we require the relation $\left\{  \Gamma_{\alpha},\gamma\right\}
=0$.

Now we consider the tunneling at the point contact. In general, there
are two major types of tunneling processes: the Majorana
fermion-induced tunneling and the regular Cooper pairs tunneling. First, we consider
the process of Majorana fermion $\gamma$ coupled to electrons of the QH liquid
$\gamma\left[  \Psi\left(  0\right)  -\Psi^{\dagger}\left(  0\right)  \right]
$. We assume the 1D TSC is sufficiently long so that Majorana fermion
$\gamma^{\prime}$ does not couple to electrons in the QH liquid. Since the
electrons in the QH liquid are fully spin-polarized (or effectively spinless),
they can couple to the Majorana mode $\gamma$ and undergo the selective
equal-spin Andreev reflection~\cite{HeJJ14PRL}, which is in analogy to the
perfect Andreev reflection in the TSC/helical Luttinger liquid
junction~\cite{FidkowskiL12PRB,Affleck13JSM}.

Second, whether the superconducting wire is in a topologically nontrivial or
trivial phase, the Cooper-pair tunneling may occur. The regular Cooper pairs may hop
from the superconducting wire into the QH liquid via the point contact, which
is induced locally in the wire by the superconducting
pairing~\cite{FidkowskiL12PRB,LeeYW14PRB,ChaoSP15PRB,Alicea11NTP}, similarly to the
regular Cooper-pair tunneling in superconductor-Luttinger liquid proximity systems~\cite{MaslovD96PRB,WinkelholzC96PRL,TakaneY97JPSJ,AffleckI00PRB,TitovM06PRL}. Since the edge states in the QH liquid can only transport spin-up electrons, the regular spin singlet Cooper-pair
tunneling process needs to complete two processes~\cite{Fisher94PRB}. The
first one is that the Cooper pair breaks apart into two unpaired electrons, as
in the conventional Andreev process, where the quantum fluctuations of
superconducting phase have been neglected. The second one is the spin-flip
process to align the spin of both electrons, which can be mediated by
spin-orbit coupling in the QH liquid. The two processes can be synthesized as
$\Psi_{\uparrow}^{\dagger}\left(  0\right)  \Psi_{\uparrow}^{\dagger}\left(
\xi\right)  +H.c,$ with $\xi$ the superconducting coherence length. If the
spin-up subscript is dropped, the tunneling Hamiltonian of Cooper pairs can be
written as $\Psi^{\dagger}\Psi^{\dagger}+H.c$. Thus, the tunneling Hamiltonian
defined at $x=0$ (the right and left sides of the junction are labelled by $x>0$ and $x<0$ respectively) reads%
\begin{equation}
H_{T}=t\gamma\left[  \Psi\left(  0\right)  -\Psi^{\dagger}\left(  0\right)
\right]  +\Delta\left[  \Psi^{\dagger}\Psi^{\dagger}+H.c\right]  ,
\end{equation}
where the first term describes the Majorana fermion tunneling with $t$
coupling amplitude and the second term accounts for regular Cooper-pair tunneling with
$\Delta$ coupling amplitude.

According to the bosonization
technology~\cite{GogolinAO98Book,GiamarchiT04Book}, the tunneling Hamiltonian
can be expressed as%
\begin{equation}
H_{T}=2it\gamma\Gamma\cos\left[  m\phi\left(  0\right)  \right]  +2\Delta
\cos2m\phi\left(  0\right)  , \label{Ht}%
\end{equation}
where the first term factorizes into the Klein-Majorana interaction and charge
sector parts. The Majorana fermion enters the problem only via $\gamma\Gamma$.
The ordinary fermion can be defined as $\psi=\left(  \gamma+i\Gamma\right)
/2$ with $\left\{  \psi,\psi^{\dagger}\right\}  =1$, so that we have%
\begin{equation}
i\gamma\Gamma=2\psi^{\dagger}\psi-1=\pm1,
\end{equation}
where $\pm$ correspond to the energy level being occupied and empty,
respectively. Consequently, the Klein-Majorana fusion procedure can eliminate
the Majorana degrees of freedom~\cite{BeriB13PRL,Alexander13PRL} and simplify
our theoretical calculations.

In what follows, we use perturbative renormalization group (RG) analysis to
uncover the transport signature of the tunneling junction. As the first step,
we integrate out the chiral bosonic field $\phi$ in the imaginary-time
partition function of the Hamiltonian in equation (\ref{Htotal}) except at
$x=0$. The partition function can be written as a path integral%
\begin{equation}
\mathcal{Z}=\int\mathcal{D}\phi e^{-(S_{0}+S_{T})},
\end{equation}
where the effective actions are given by
\begin{equation}
S_{0}=\frac{m}{2\pi}\int\frac{d\omega}{2\pi}\left\vert \omega\right\vert
\left\vert \phi(\omega)\right\vert ^{2},
\end{equation}
and%
\begin{equation}
S_{T}=\int d\tau\left[  2t\cos\left[  m\phi\left(  \tau\right)  \right]
+2\Delta\cos2m\phi\left(  \tau\right)  \right]  .
\end{equation}

Next, we perform weak-coupling RG analysis in the frequency domain. The
bosonic field $\phi$ is separated into slow ($s$) and fast ($f$) modes:
$\phi_{s}\left(  \tau\right)  =\int_{-\Lambda/b}^{\Lambda/b}\frac{d\omega
}{2\pi}e^{-i\omega\tau}\phi(\omega)$ and $\phi_{f}\left(  \tau\right)
=\int_{\Lambda/b<\left\vert \omega\right\vert <\Lambda}\frac{d\omega}{2\pi
}e^{-i\omega\tau}\phi(\omega)$, with $\Lambda$ as an energy cutoff, $b>1$ as a
scale factor, and $\tau=it$ the Euclidean time.

After integrating over the fast modes, the new effective action can be
calculated using cumulant expansion to the lowest-order approximation in
coupling $t$ and $\Delta$, yielding
\begin{equation}
e^{-S_{eff}\left[  \phi_{s}\right]  }\approx e^{-S_{s}\left[  \phi_{s}\right]
}e^{-\left\langle S_{T}\left[  \phi_{s},\phi_{f}\right]  \right\rangle _{f}},
\end{equation}
where $\left\langle \ldots\right\rangle _{f}$ denotes integrating out the fast
mode, and $S_{s}\left[  \phi_{s}\right]  $ is the slowly fluctuating part of
$S$. The lowest-order flows of the coupling $t$ and $\Delta$ under the RG are
given by%
\begin{eqnarray}
\frac{dt}{d\ln b}  &  =t\left(  1-\frac{m}{2}\right)  ,\label{RGFQH}\\
\frac{d\Delta}{d\ln b}  &  =\Delta\left(  1-2m\right)  , \label{RG-Cooper}%
\end{eqnarray}
with $\ln b$ as the dimensionless flow parameter. For the $p$-wave Cooper-pair
tunneling, we have $\Delta\left(  \Psi^{\dagger}\partial_{x}\Psi^{\dagger
}+H.c\right)  \propto\sin2m\phi$, from which the same lowest-order flow
equation as equation (\ref{RG-Cooper}) can be obtained. In addition, we may in
principle add other types of coupling at the point contact, including even
higher-order derivatives in $\Psi^{\dagger}\left(  0\right)  $ and/or
$\Psi\left(  0\right)  $, but it can be shown that they are generally
irrelevant~\cite{FidkowskiL12PRB,Affleck13JSM}.

From these flow equations (\ref{RGFQH}) and (\ref{RG-Cooper}), we can reach
the following conclusions. For all single-channel QH liquid, the Cooper-pair
tunneling is irrelevant. For the integer QH liquid of $m=1$, the tunneling
process of Majorana fermions coupling to electrons of the QH liquid is
relevant. However, for the Laughlin fractional QH liquid of $m\geq3$, the
Majorana fermions tunneling is irrelevant. So, for 1D TSC/integer QH liquid
tunnel junction, the perfect normal reflection fixed point is unstable toward
the perfect Andreev reflection fixed point, which is analogues to the hybrid
system of superconductor trench and QH liquid\cite{ClarkeDJ14NTP}. In what
follows, we first analyze the stability of the perfect Andreev reflection
fixed point and then the transport properties of 1D TSC/fractional QH liquid
tunnel junction.

\subsection{Integer QH liquid}

In the case of $m=1$, one sees that the tunneling term $t$ is relevant. The
universal low-energy transport is governed by the fixed point describing
perfect Andreev reflection with quantized zero-bias conductance $G(0)=2e^{2}%
/h$. At this fixed point, the nontrivial boundary condition $\Psi^{\dagger}\left(
0^{+}\right)  =e^{i\alpha}\Psi\left(  0^{-}\right)  $ ($0^{\pm}$ stands for the
right and left sides of the point contact junction) needs to be
satisfied~\cite{FidkowskiL12PRB,ClarkeDJ14NTP}. Here, we set $\alpha=0$ for
simplicity and this boundary condition implies $\phi\left(  0^{+}\right)
=-\phi\left(  0^{-}\right)  $. Next, we investigate the stability of the
perfect Andreev reflection fixed point. Physically, the deviations from this
fixed point mean that the electrons $\psi\left(  x\right)  =e^{i\phi\left(
x\right)  }$(we have neglected the Klein factor. For the fractional QH liquid,
the $e^{i\phi\left(x\right)}$ is fractionally-charged quasiparticle) transmission between left and right edge parts of QH
liquid at $x=0$ is allowed. So, the leading perturbation away from the perfect
Andreev reflection fixed point is the electrons (fractionally-charged quasiparticles for the
fractional QH liquid) transmission along the edge of the QH liquid at $x=0$
\begin{equation}
\delta S_{u}=u\int d\tau\left[  \psi^{\dagger}\left(  0^{+}\right)
\psi\left(  0^{-}\right)  +H.c\right].
\end{equation}

According to the RG transformation, the lowest-order flow of coupling $u$ can
be expressed as%
\begin{equation}
\frac{du}{d\ln b}=u\left(  1-\frac{2}{m}\right)  , \label{uAndreev}%
\end{equation}
from which, it follows that for the integer QH liquid ($m=1$), the
electrons transmission is irrelevant. As a result, the perfect Andreev
reflection fixed point due to the Majorana mode coupling is stable and the
resulting ZBCP in tunneling spectra is robust. For the fractional QH liquid,
the quasiparticles transmission is relevant and the perfect Andreev reflection fixed
point is unstable.

Based on the analysis above, we can conclude that for the 1D TSC/integer QH
liquid tunnel junction, there is a perfect Andreev reflection fixed point with
zero-bias tunneling conductance $G(0)=2e^{2}/h$ at zero temperature and
voltage. Such a quantized zero-bias tunneling conductance is robustly
independent of those irrelevant parameters, so long as the coupling of the
Majorana modes to the integral QH liquid remains finite. In contrast, for the
topologically trivial superconductor ($s$-wave or $p$-wave) in the absence of
a Majorana fermion, the ground state corresponds to a perfect normal
reflection fixed point with a vanishing zero-bias tunneling conductance
$G(0)=0$. From the distinguishing difference in zero-bias tunneling
conductance between the superconductors in the presence and absence of
Majorana modes, it follows that the quantized conductance $G(0)=2e^{2}/h$ can
serve as a definitive fingerprint of a Majorana fermion at the 1D TSC/integer
QH liquid tunnel junction, and the ZBCP can be regarded as an illustrative
example of Majorana fermion-induced equal-spin Andreev reflections. Compared
to the normal metal
lead\cite{LawKT09PRL,FidkowskiL12PRB,Affleck13JSM,KomijaniY14PRB}, there are
outstanding advantages of integer QH liquid lead. The electronic interactions
in the integer QH liquid are suppressed. The disorder plays a decisive role in
the formation of the observed integer QH plateaus~\cite{Yoshioka02Book}. So,
the effects of various factors such as electronic interactions, disorder, and
finite length on the tunneling spectrum are removed.

\subsection{Fractional QH liquids}

From the equations (\ref{RGFQH}) and (\ref{RG-Cooper}), we can see that for
the Laughlin fractional QH liquid, the Majorana fermions and Cooper-pair
tunneling are irrelevant. The perfect normal reflection fixed point is stable.
From equation (\ref{uAndreev}), it is found that the quasiparticles transmission is
relevant and the perfect Andreev reflection fixed point is unstable. So, no
matter that the superconducting wire is nontrivial or trivial, the low-energy
physics for all Laughlin fractional QH liquids is governed by the perfect
normal reflection fixed point with a vanishing zero-bias conductance, $G(0)=0$.

Based on the analysis above, when the superconducting wire is nontrivial, the
leading-order perturbation around the perfect normal reflection fixed point is
the Majorana fermion-induced tunneling. To make a connection with experiments,
it is important to take into account the corrections to the tunneling
conductance due to finite bias voltage or temperature. According to the equation (\ref{RGFQH}), we can calculate the
tunneling conductance to the lowest-order approximation in bias voltage $V$ or
temperature $T$, yielding%
\begin{equation}
G\left(  V\right)  \sim V^{m-2},G\left(  T\right)  \sim T^{m-2}.
\label{TunnelConduct1}%
\end{equation}

This power law vanishing form can be understood by the following argument.
First, the quasiparticles in the Laughlin fractional QH liquid are anyons with
fractional charge and fractional exchange statistics, which are not allowed
for tunneling from the Laughlin fractional QH liquid to the TSC. These anyons
must form into electrons for tunneling. Second, the Majorana fermion
(Majorana bound state) is the Bogoliubov quasiparticle with equal
superposition of a electron and a hole. On the other hand, fractional QH edge
states are topological many-body systems with strong repulsive interaction, which
make the Majorana fermions tunneling strongly inhibited in contrast to
integer QH liquid case. Third, the fractional QH liquid is phase coherence
state and does not dissipate, so the Majorana fermions of TSC do
not reach equilibrium with the reservoir of edge excitations of the fractional QH
liquid. In other words, the physics of Majorana fermion-induced tunneling is
determined by the physical properties of quasiparticles in the fractional QH liquid
and Majorana fermions in TSC. All these factors make the point contact insulating.

For the trivial superconducting wire, the leading-order perturbation around
the perfect normal reflection fixed point is the regular Cooper-pair tunneling due to
absence of Majorana fermions. So, for a Laughlin QH liquid coupled to a
topologically trivial $s$-wave ($p$-wave) superconductor, the tunneling
conductance is given by~\cite{Fisher94PRB}%
\begin{equation}
G\left(  V\right)  \sim V^{4m-2},G\left(  T\right)  \sim T^{4m-2}%
,\label{TunnelConduct2}%
\end{equation}
which is also applicable to the integral QH liquid case. As discussed above,
the regular Copper-pair tunneling is composite process. The Cooper pair first breaks
into two unpaired electrons. Then, the spin-down electron is changed into
a spin-up electron by the spin-flip process. Because of the Pauli exclusion
principle, these two spin-up electrons can not occupy the same state
(position). One electron needs to get away some certain distance for the
other to transfer. For the $p$-wave Cooper-pair tunneling, there is no spin-flip
process but the Pauli exclusion principle still plays a role on the tunneling.
On the basis of the strongly repulsive Laughlin fractional QH liquid, the
tunneling of the $s$-wave and $p$-wave Cooper pairs is strongly inhibited at
low energies because of the Coulomb blockade effect and Pauli exclusion principle.

A comparison between equations (\ref{TunnelConduct1}) and
(\ref{TunnelConduct2}) indicates that although the zero-bias tunneling
conductances are vanishing in both cases, they exhibit quite different
power-law scaling with both bias and temperature for fixed $m$, from which one
can judge whether the Majorana fermion is present or absent in the
superconducting wire. Take the $\nu=1/3$ Laughlin fractional QH liquid
as example. For the 1D TSC, the tunneling conductance varies linearly
with bias voltage or temperature at low energies, while for the conventional
superconductor, it is proportional to $V^{10}$ (or $T^{10}$). From
the experimental observation of tunneling conductance, it is easy to distinguish
its linear scaling with $V$ from the tenth power of $V$, providing an
indication of the existence of the Majorana modes in the 1D TSC. On the other
hand, from the tunneling conductance behavior, we can extract the topological
invariant $m$ controlled by the bulk fractional QH states, whose value does
not depend on detailed properties of edges. Similar to the integer QH liquid
case, the electronic interactions in the fractional QH liquid are fixed. The
fractional QH states are robust to impurity such as disorder. Thus, the
TSC/fractional QH liquid tunnel junction has the same outstanding advantages
of the integer QH liquid case.

The transport measurements of our setup are within reach of present
experimental techniques. The 1D spin-orbit coupled wire can be realized using
an InSb nanowire with effective mass $m^{\ast}=0.015m_{e}$~\cite{Mourik12SCI}.
The $g$ factor is equal to $g=50,$ the magnetic field $B\sim0.15T$, and the
Rashba spin-orbit coupled $E_{SO}\sim50\mu eV$. Its proximity to
superconducting NbTiN electrodes can induce a superconducting gap of about
$250\mu eV$. On the other hand, the QH liquid can be obtained by GaAs-AlGaAs
heterojunctions~\cite{Tsui82PRL}. However, the QH liquids need high magnetic
field, which destroys the superconductivity of the s-wave superconductor in TSC.
We can use a type-II s-wave superconductor in TSC compatible with the high magnetic
field in QH liquids. In addition, the $\nu=1$ integer QH liquid
can be achieved by the quantum anomalous Hall states in magnetic
topological insulators~\cite{ChangCZ13SCI}, which do not need the high magnetic
field. Furthermore, the quantum point contacts have been used widely as transport measurements in solid-state systems.

\section{Conclusions}

\begin{table}[!htbp]
\caption{The tunneling conductances at finite voltage and zero temperature of
four tunnel junction archetypes. The columns denote two types of
single-channel QH liquids, and the rows indicate the superconductors (SC) with
and without Majorana zero-mode bound states.}
\label{table1}
\begin{tabular}{lll}
\hline
\phantom{blah}& integral QH & fractional QH\\
\hline
topological SC & $\frac{2e^{2}}{h}(1-c_{V}V^2)^{*}$ & $\sim V^{m-2}$ \\
conventional SC & $\sim V^2$ & $\sim V^{4m-2}$ \\
\hline
\end{tabular}
\begin{tablenotes}
\item[*] $c_{V}$ denotes a non-universal constant
\end{tablenotes}
\end{table}

The point contact tunnel junctions between a 1D TSC and single-channel QH
liquids have been investigated by bosonization technology and renormalization
group methods. The main results are partially reported in table \ref{table1}.
Due to different correlated and topological properties, the tunneling
conductance behaviors of four types of junctions are quite different from each
other. For a 1D TSC coupled to a $\nu=1$ integer QH liquid, the low-energy
physics is governed by the perfect Andreev reflection fixed point. We have
proposed an experimentally accessible scheme with relatively simple geometry
to detect the ZBCP, the tunneling being induced by the Majorana modes rather
than by the conventional Cooper pairs. As a result, the quantized zero-bias
conductance $2e^{2}/h$ can serve as a definitive fingerprint of the existence
of Majorana fermions. If the Majorana fermion is absence, we have $G(V)\sim
V^{2}$, exhibiting a vanishing zero-bias conductance. In the $\nu=1/m$
Laughlin QH liquid case, the universal low-energy transport is governed by the
fixed point describing perfect normal reflection, and the bias-dependent
tunneling conductance is given by $G(V)\sim V^{m-2}$ and $V^{4m-2}$,
respectively, for the 1D TSC and conventional SC. The temperature and voltage
dependence of the tunneling conductance reflect the topological structure of
the fractional QH liquids. From the behavior of tunneling conductance, we can
extract the topological invariant $m$ controlled by the bulk fractional QH
states. An outstanding advantage of using the QH liquids is that only two
factors, the topological character of the TSC and filling $\nu$ of the QH
liquid, determine the characteristic of tunneling conductance at low energy,
the others such as electronic interaction strength, disorder, and finite
length of the lead are irrelevant. There are some open works for future research. We can use the BTK theory~\cite{BTK82PRB} with scaling approach and entire S-matrix~\cite{TitovM06PRL} to recalculate the tunneling transport properties and discuss the effects of geometry, boundaries, multi-band, and disorder of the superconducting wire and incoherent reservoir. A numerical study of our system would be very useful, checking our predicted phase diagram.

\acknowledgments
This work was supported by the National Natural Science Foundation of China
under grant numbers 11447008 (Z.Z.W.), 11225420 (L.S.),and 11474149 (R.S.),
the State Key Program for Basic Researches of China under grants numbers
2014CB921103 (L.S.) and 2011CB922103 (D.Y.X.), and a project funded by the
PAPD of Jiangsu Higher Education Institutions.


\end{document}